\documentclass[prx,twocolumn,preprintnumbers,superscriptaddress,longbibliography, nofootinbib, 10pt]{revtex4-2} 
\usepackage{graphicx}
\usepackage{bm}
\usepackage{amsmath}
\usepackage{amssymb}
\usepackage{amsfonts}
\usepackage{slashed}
\usepackage{latexsym,epsfig,bbm}
\usepackage{mathtools}
\usepackage{color}
\usepackage{algorithm}
\usepackage{algorithmic}
\definecolor{red}{rgb}{0.9,0,0}

\usepackage{hyperref}
\usepackage{amsthm}
\usepackage{braket}				               

\newcommand{\past}      { \smash{\overleftarrow {\meassymbol}} }

\newcommand{\CausalState}       { \mathcal{S} }

\newcommand{\forward}{+}
\newcommand{\reverse}{-}
\newcommand{\forwardreverse}{\pm} 

\newcommand{\FutureCausalState} { {\CausalState}^{\forward} }

\newcommand{\PastCausalState}   { {\CausalState}^{\reverse} }

\newcommand{\lastindex}[2]{
  \edef\tempa{0}
  \edef\tempb{#2}
  \ifx\tempa\tempb
    \edef\tempc{#1}
  \else
    \edef\tempa{0}
    \edef\tempb{#1}
    \ifx\tempa\tempb
      \edef\tempc{#2}
    \else
      \edef\tempc{#1+#2}
    \fi
  \fi
  \tempc
}

\newcommand{\CSjoint}[1][,]{
   \edef\tempa{:}
   \edef\tempb{#1}
   \ifx\tempa\tempb
      \ensuremath{\FutureCausalState\!#1\PastCausalState}
   \else
      \ensuremath{\FutureCausalState#1\PastCausalState}
   \fi
}
\newcommand{\CSjointKL}[3][,]{
   \edef\tempa{:}
   \edef\tempb{#1}
   \ifx\tempa\tempb
      \ensuremath{\FutureCausalState_{#2}\!#1\PastCausalState_{#3}}
   \else
      \ensuremath{\FutureCausalState_{#2}#1\PastCausalState_{#3}}
   \fi
}

\newif\ifpm
\edef\tempa{\forwardreverse}
\edef\tempb{\pm}
\ifx\tempa\tempb
   \pmfalse
\else
   \pmtrue
\fi

\renewcommand{\past}       {\ms{}{0}}

\ifcsname mathclap\endcsname
  \relax
\else
  \def\clap#1{\hbox to 0pt{\hss#1\hss}}

\fi

\newcommand{\op} [3] [] {
  \ensuremath{
    \operatorname{#2_{#1}}
    \if\relax\detokenize{#3}\relax
    \else
      \left[ #3 \right]
    \fi
  }
  \xspace
}

\theoremstyle{plain}    
\theoremstyle{plain}    
\theoremstyle{plain}    \newtheorem{Cor}{Corollary}
\theoremstyle{plain}    
\theoremstyle{plain}    
\theoremstyle{plain}    
\theoremstyle{plain}    \newtheorem{Prop}{Proposition}
\theoremstyle{plain}    
\theoremstyle{plain}    
\theoremstyle{plain}    
\theoremstyle{plain}    \newtheorem{Def}{Definition}
\theoremstyle{plain}

\newcommand{\SSet}{\boldsymbol{\mathcal{S}}}

\newcommand{\stationary}{\boldsymbol{\pi}}

\newcommand{\tr}{\text{tr}}
\newtheorem{thm}{Theorem}
\newcommand{\model}{\boldsymbol{\mathcal{M}}}
\newcommand{\modelq}{{\boldsymbol{\mathcal{M}}_q}}
\newcommand{\futureWL}{\Omega_\text{f}}
\newcommand{\pastWL}{\Omega_\text{h}}

\newcommand{\sysref}{I/d}  
\newcommand{\constname}{\xi}

\newcommand{\ones}{\boldsymbol{1}}
\renewcommand{\past}[1]{\overleftarrow{#1}}
\newcommand{\fut}[1]{\overrightarrow{#1}}

\begin{document}

\title{Identifiability and minimality bounds of quantum and post-quantum\\
models of classical stochastic processes}

\author{Paul M.\ Riechers}
\email{pmriechers@gmail.com}
\affiliation{Beyond Institute for Theoretical Science (BITS), San Francisco, California}
\author{Thomas J.\ Elliott}
\email{physics@tjelliott.net}
\affiliation{Department of Physics \& Astronomy, University of Manchester, Manchester M13 9PL, United Kingdom}
\affiliation{Department of Mathematics, University of Manchester, Manchester M13 9PL, United Kingdom}
\affiliation{Centre for Quantum Science and Engineering, University of Manchester, Manchester M13 9PL, United Kingdom}

\date{\today}

\begin{abstract}
To make sense of the world around us, we develop models, constructed to enable us to replicate, describe, and explain the behaviours we see. Focusing on the broad case of sequences of correlated random variables, i.e., classical stochastic processes, we tackle the question of determining whether or not two different models produce the same observable behavior. This is the problem of identifiability. Curiously, the physics of the model need not correspond to the physics of the observations; recent work has shown that it is even advantageous---in terms of memory and thermal efficiency---to employ quantum models to generate classical stochastic processes. We resolve the identifiability problem in this regime, providing a means to compare any two models of a classical process, be the models classical, quantum, or `post-quantum', by mapping them to a canonical `generalized' hidden Markov model. Further, this enables us to place (sometimes tight) bounds on the minimal dimension required of a quantum model to generate a given classical stochastic process.
\end{abstract}

\date{\today}
\maketitle

\section{Introduction}

Modern science is built upon \emph{models}~\cite{burnham2003model}. From the observations we make, we hypothesize underlying mechanisms that could give rise to them, and test these proposed models by making predictions about as-yet-unseen observations. Many---in fact, an infinite number---of different models can give rise to the exact same observable behaviors, motivating the pursuit for the most parsimonious models of a process. A natural question is that of how we determine these minimal models.

One route is through information theory. Mechanistically, physical systems carry information in their state as an intrinisic memory of the past features that give rise to future behaviors. How much internal structure, though, is required to generate complex observable output?~\cite{Crut92c} We can devise models that obey different physical laws than the observable behaviors they generate, and perhaps surprisingly, this can often be advantageous~\cite{Gu12a}. Quantum mechanics allows for physical systems to exhibit observable classical behaviors whilst bearing less memory than would be required of any classical generator~\cite{elliott2021memory}. 

In addition to lower entropy, quantum models can allow significant dimensional advantage, which we focus on here.
That is, if a quantum state is used as the memory for the generator, then the dimension of the quantum state can be much smaller than the minimal number of classical memory states that would be required to produce the same process~\cite{Monras16_Quantum}. In other words, quantum explanations of classical behavior are often simpler.

Faced with such a broad spectrum of potential models---classical or quantum---of processes, how do we determine when two seemingly-disparate models generate the same observable behaviors? This is the \emph{identifiability problem}~\cite{Blac57a}. Here, we resolve the identifiability problem in a unified manner for classical, quantum, and even more general `post-quantum' linear models. We do so by mapping all such mechanistic descriptions to a minimal, canonical linear latent model. As a byproduct, we lower-bound the minimal number of quantum dimensions necessary to generate any particular classical stochastic process.

Indeed, directly encoding the physical constraints of classical microstates or quantum density operators into a search over such models results in a challenging constrained optimization problem. The relaxation of the search space to more general linear models---of which the classical and quantum are a subset---simplifies the problem considerably, allowing us to identify a unique, minimal generative model of a process within this space. We begin in Sec.~\ref{sec:framework} by introducing these general linear models over latent variables, termed generalized hidden Markov models (GHMMs)~\cite{Uppe97a}---which naturally represent generalized probabilistic theories that include classical and quantum frameworks as special cases. Sec.~\ref{sec:framework} also introduces 
our notation and core concepts. We then show in Sec.~\ref{sec:Identifiability} how the uniqueness of the minimal GHMM enables us to solve the identifiability problem, and moreover, set bounds on classical and quantum generative models in Sec.~\ref{sec:Minimizing}.

\section{Framework}
\label{sec:framework}

\subsection{Stochastic Processes}
\label{sec:stochproc}

A discrete classical \emph{stochastic process} consists of a sequence of random events $X_l$, taking observable values $x_l\in\mathcal{X}$, with the index $l$ indicating their position in the sequence~\cite{khintchine1934korrelationstheorie}. These events
can be intricately correlated, and are drawn consistently from the joint probability distribution over sequential events.  We express contiguous marginal distributions over subsequences as 
$P(X_{l:l'})$, where we use the shorthand $X_{l:l'}:=X_l X_{l+1}\ldots X_{l'-1}$ to denote a contiguous sequence of random variables\footnote{In this notation, the left index is inclusive, the right exclusive.}.

When considering stationary stochastic processes, such that the position index can take any integer value, and the distributions are shift invariant, i.e., $P(X_{l:l'})=P(X_{l+L:l'+L})\forall l,l',L\in\mathbb{Z}$ such that $l' > l$ and $L\geq0$, we can neglect absolute references to the positions of events in a sequence.
For shorthand, we will use $w\in\mathcal{X}^{|w|}$ to represent a consecutive sequence of events (a \emph{word}) of length $|w|$. For example, with $w=x_{0:L}$ and $w'=x_{L:L'}$ with $L'>L$, we can write the conditional probability of $w'$ given $w$ as $P(w'|w):=P(X_{L:L'}=w'|X_{0:L}=w)$. The contiguous concatenation of words $w$ and $w'$ is simply written as $ww'$.

The stochastic processes we consider here are classical, such that the distributions obey the Kolmogorov consistency conditions~\cite{kolmogorov2018foundations}, and may be in principle generated by a system with a---potentially infinite---classical state space, undergoing classical dynamics. Nevertheless, the state space and dynamics of the system that generates such a process may more generally be classical, quantum, or even belong to the broader class of generalized probabilistic theories~\cite{garner2015phase}.

\subsection{HMMs and GHMMs}
\label{sec:GHMMs}

\begin{figure}
	\includegraphics[width=\columnwidth]{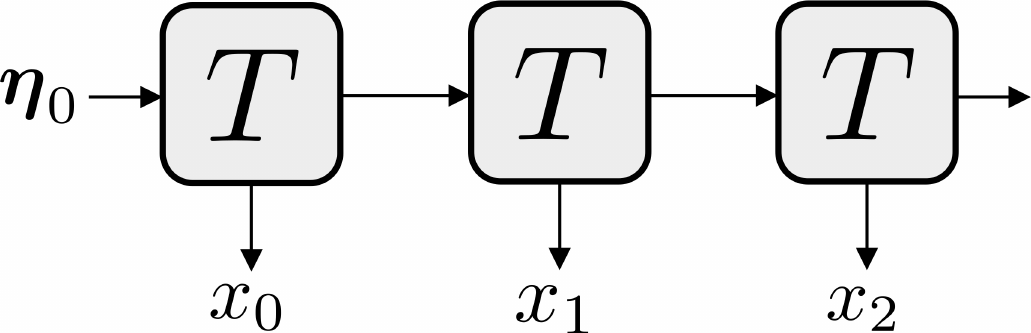}
	\caption{A classical stochastic process can be generated by a [generalized] hidden Markov model ([G]HMM). A GHMM consists of a series of transition operators $\{T^{(x)}\}$ and initial state $\boldsymbol{\eta}_0$, such that iterated application of the transition operators on the initial state generates a correlated sequence of random variables $X_0X_1X_2\ldots$, corresponding to the process.}
	\label{fig:ghmm}
\end{figure}

The state space, dynamics, and observations from a generalized probabilistic theory generating a discrete classical stochastic process, including classical and quantum processes as special cases, can be represented by \emph{generalized hidden Markov models} (GHMMs) [see Fig.~\ref{fig:ghmm}]. GHMMs, introduced by Upper in Ref.~\cite{Uppe97a}, have nearly-equivalent representations sometimes known as `quasi-realizations'~\cite{Monras16_Quantum, Fanizza24_Quantum} or `weighted automata'~\cite{Balle15_Canonical}.

A GHMM is defined by the three-tuple
\begin{align}
	\model = \bigl( \mathcal{X} ,  \boldsymbol{\eta}_0 , (T^{(x)})_{x \in \mathcal{X}} \bigr) ,
\end{align}	
where $\mathcal{X}$ is the set of observable events, $\boldsymbol{\eta}_0$ is the initial vector in a (real) vector space, and the net transition operator $T = \sum_{x \in \mathcal{X}} T^{(x)}$ has an eigenvalue of unity with associated right eigenvector $\ones = T \ones$, normalized such that $\boldsymbol{\eta}_0 \boldsymbol{1} = 1$. The associated left eigenvector $\stationary = \stationary T$, also normalized such that $\stationary \ones=1$, is the \emph{steady-state} of the GHMM.

The probability of any sequence $w = x_{0:L} \in \mathcal{X}^L$ of arbitrary length $L$ can be calculated directly via linear algebra:
\begin{align}
	P_{\model}(X_{0:L} = w) =  \boldsymbol{\eta}_0 T^{(w)} \ones  ~,
	\label{eq:Probs_from_GHMM}
\end{align}	
where $T^{(w)} = T^{(x_0)} \cdots T^{(x_{L-1})} $. At any $L$, normalization in probability follows from 
\begin{align}
	\sum_{w \in \mathcal{X}^L }	P_{\model}(w) 
	=  \sum_{w \in \mathcal{X}^L } \boldsymbol{\eta}_0 T^{(w)} \ones 
	= \boldsymbol{\eta}_0  T^L  \ones  
	=    \boldsymbol{\eta}_0 \ones
	= 1.
\end{align}	
A GHMM is \emph{valid} if and only if it produces non-negative probabilities for each word, i.e., $ \boldsymbol{\eta}_0 T^{(w)} \ones  \geq 0\forall w \in \mathcal{X}^*$.

\emph{Hidden Markov models}\footnote{Specifically, Mealy HMMs, which produce observables from edges when transitioning between hidden states.  The alternative, Moore HMMs, produce observables according to probability distributions associated with each hidden state.  They are equivalent in the sense that if a stochastic process is finitely-representable by one, then it is finitely-representable by the other.} (HMMs) are a special case of GHMMs with non-negative transition elements: \mbox{$T^{(x)}_{s,s'}\geq 0$}. The transition operators $\{T^{(x)}\}$ of a HMM are substochastic, with the net transition operator \mbox{$T = \sum_{x \in \mathcal{X}} T^{(x)}$} row-stochastic. $T$ has an eigenvalue of unity, with the corresponding (normalized) right eigenstate $\ones = T \ones$ a column vector of all ones, and the corresponding 
left eigenstate $\stationary = \stationary T$, normalized such that $\stationary \ones = 1$, a stationary distribution over hidden states of the model. A HMM is \emph{unifilar} if there is at most one non-zero element in each column of each $T^{(x)}$\footnote{A more conceptual picture of unifilarity is that if given any start state with a single non-zero element (i.e., there is no uncertainty in the initial state), then given any particular output word $w$ the final state also has only one non-zero element (i.e., there is no uncertainty in the final state). This mirrors the concept of determinism for finite automata~\cite{Shal98a}.}.

\subsection{QHMMs}

Classical stochastic processes can also be generated by quantum models. Such \emph{quantum hidden Markov models} (QHMMs) operate through the action of a quantum operation on a quantum `memory' system~\cite{monras2010hidden}. Classical HMMs constitute a special case of QHMMs; any classical stochastic process that can be generated by a finite-state HMM can be generated by a finite-dimensional QHMM~\cite{glasser2019expressive, elliott2021memory}. 

Formally, a QHMM can be specified by the tuple 
\begin{equation}
\label{eq:qhmmdef}
\modelq = (\mathcal{X}, \sigma_0, (A_x)_{x\in\mathcal{X}}),
\end{equation}
where $\mathcal{X}$ is the set of observable events, $\sigma_0$ is the initial state of the QHMM memory system (corresponding to a density operator on a $d_{\modelq}$-dimensional Hilbert space), and $(A_x)_{x\in\mathcal{X}}$ is a collection of trace non-increasing linear superoperators that act on the QHMM memory system, playing an analogous role to the transition operators of a (G)HMM. These superoperators have a representation in terms of Kraus operators $(K_{xy})_{x\in\mathcal{X},y\in\mathcal{Y}}$ acting on the memory system, where $\mathcal{Y}$ is an alphabet of unobserved `trashed' events:
\begin{equation}
A_x(\rho) =  \sum_{y \in \mathcal{Y}}K_{xy} \rho K_{xy}^\dagger,
\end{equation}	
where $\sum_{x \in \mathcal{X}, y \in \mathcal{Y}} K_{xy}^\dagger K_{xy} = I$. We remark that in prior literature QHMMs are typically specified in terms of the Kraus operators $\{K_{xy}\}$~\cite{monras2010hidden, Monras16_Quantum}, or a unitary operator that induces them~\cite{binder2018practical, liu2019optimal, elliott2021memory}, rather than in terms of the associated superoperators. However, we define QHMMs in this manner in order to make the connection with GHMMs more readily apparent, which we lean on in proofs below. All these representations are functionally equivalent.

We can construct operators corresponding to words of consecutive outputs by consecutive action of the superoperators. For example, for $w=x_{0:L}$ we have 
\begin{equation}
	A_w(\cdot) = A_{x_{L-1}} \bigl( A_{x_{L-2}} \bigl( \cdots A_{x_{0}} \bigl( \cdot \bigr) \ldots \bigr) \bigr).
\end{equation}
The corresponding word probabilities are then given by
\begin{equation}
	P_{\modelq}(w) = \tr \Bigl[  A_w(\sigma_0) \Bigr].
\end{equation}

Furthermore, we define the normalizing operation $N$ which acts as 
\begin{align}
	N(\rho) = \rho / \tr(\rho).
\end{align}	
on operators $\rho$ of non-zero trace. This allows us to define an update rule for the memory conditional on the observation; upon observation $x_{l}$, the memory system state updates as:
\begin{align}
	\sigma_{l+1} = N( A_{x_{l}} (\sigma_l)).
\end{align}	
We can also use this to specify conditional probabilities:
\begin{align}
	P_{\modelq}(w' | w) = \tr \Bigl[ A_{w'} \bigl( N[A_w(\sigma_0)] \bigr) \Bigr].
\end{align}	

\begin{figure}
	\includegraphics[width=\columnwidth]{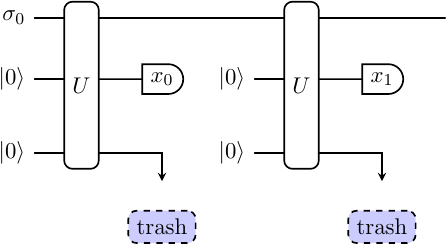}
	\caption{We address identifiability and minimality of quantum models of classical stochastic processes. A QHMM, as depicted in the figure, generates a classical stochastic process $X_1X_2\ldots$	by repeated application of a joint unitary transformation $U$ acting on a quantum memory (initialized in state $\sigma_0$), a blank state $\ket{0}$ for the subsequently observed event, and (possibly) an auxiliary system, also initialized in state $\ket{0}$, that is trashed following the transformation. QHMMs generalize HMMs, with those without the trashed auxiliary system providing a quantum generalization of unifilar HMMs.}
	\label{fig:UnitaryGenerator}
\end{figure}

Mechanistically, a QHMM can be realized by the action of a joint unitary transformation $U$ 
on a supersystem composed of the quantum memory and an ancilla.
While the memory system has a $d_{\modelq}$-dimensional Hilbert space,
the ancillary system has a Hilbert space of dimension $|\mathcal{X}|\times|\mathcal{Y}|$ prepared in a blank fiducial state $\ket{0}\ket{0}$. Measurement of the first subsystem of the ancilla then yields the output event for a given timestep, and the second subsystem is trashed\footnote{In the case of $|\mathcal{Y}|=1$, the second subsystem may be omitted. Under such circumstances, the QHMM is unifilar. Unifilarity in a quantum model can be seen conceptually in terms of any initial pure state of the memory and specified output word leading to a final pure state.}. Repeated iteration of this procedure generates the full process. This is depicted in Fig.~\ref{fig:UnitaryGenerator}. We can relate $U$ to the superoperators $\{A_x\}$ via
\begin{align}
	A_x(\rho) \! &= \! \sum_{y \in \mathcal{Y}} \! (I \!\otimes\!\bra{x} \! \bra{y})  U \bigl( \rho \otimes \ket{0} \!\! \bra{0} \otimes \ket{0} \!\! \bra{0} \bigr) U^\dagger (I \! \otimes \!\ket{x} \! \ket{y} \!).
\end{align}	
Correspondingly, we can specify a QHMM in terms of $U$ rather than the superoperators $(A_x)_{x\in\mathcal{X}}$. A given $U$ uniquely specifies $(A_x)_{x\in\mathcal{X}}$ through the above relation, while, much like quantum state purification, a given set of superoperators $(A_x)_{x\in\mathcal{X}}$ admits infinitely many dilations into unitary operators.

Above, we noted that HMMs are a special case of QHMMs. In fact, the set of classical stochastic processes that can be generated by a finite HMM is a strict subset of those that can be generated by a finite QHMM~\cite{adhikary2020expressiveness, Fanizza24_Quantum}. Prior literature has further established that even amongst processes that can be generated by finite HMMs, the memory entropy and dimension of the minimal QHMM can be strictly lower than that of the minimal HMM, in both the case of unifilar~\cite{Gu12a, loomis2019strong, liu2019optimal, ghafari2019dimensional, elliott2020extreme} and non-unifilar generators~\cite{Monras16_Quantum, elliott2021memory, Fanizza24_Quantum, zonnios2025quantum}. Similar advantages can be found in terms of the thermodynamic cost of process generation~\cite{loomis2020thermal, elliott2021memory}. Conversely, there exist classical stochastic processes that can be generated by finite GHMMs, but not by a finite QHMM~\cite{Fanizza24_Quantum}. Nevertheless, as we demonstrate in an example below, some processes can be generated by QHMMs with a smaller memory dimension than that of the minimal GHMM.~\footnote{We emphasize that multiple notions of dimension are at play here.  The physical dimension, i.e., the Hilbert space dimension, of quantum memory may be smaller than the latent dimension of the smallest GHMM.  Yet, we note that the dimension of the vector space spanned by the memory states (as quantum mixed states) can never be less than the latent dimension of the smallest GHMM.}

\section{When are two models equivalent?}
\label{sec:Identifiability}

When do two different models---be they a classical HMM, a quantum QHMM or a post-quantum GHMM---generate the same classical stochastic process? This is the identifiability problem, which asks when two generative statistical models (e.g., two different HMMs) represent the same process~\cite{Blac57a}. This problem has been addressed for HMMs and GHMMs---see, for example, Thm.\ 4.3.3 of Ref.~\cite{Uppe97a}. Following a closely analogous approach, we extend this result to further incorporate QHMMs. We do this by mapping QHMMs to a canonical `standard' GHMM presentation. This first requires that we define the following spaces, derived from the QHMM.

\begin{Def}
The \emph{history space} $\mathcal{H}_{\modelq}$ induced by quantum model $\modelq$ is the span of observation-induced states of the quantum memory:
\begin{align}
	\mathcal{H}_{\modelq} \coloneqq \text{\emph{span}} \bigl\{ A_w(\sigma_0) : w \in \mathcal{X}^* \bigr\}.
\end{align}
\end{Def}

\begin{Def}
	The \emph{future space} $\mathcal{F}_{\modelq}$ induced by quantum model $\modelq$ is the span of the linear functionals induced by future words:
	\begin{align}
		\mathcal{F}_{\modelq} \coloneqq \text{\emph{span}} \bigl\{ \text{\emph{tr}} \bigl[ A_w( \cdot ) \bigr] : w \in \mathcal{X}^* \bigr\}.
	\end{align}
\end{Def}

Note that $\varnothing \in \mathcal{X}^*$, which induces the identity map $A_\varnothing = I$.  Accordingly, $\sigma_0 \in \mathcal{H}_{\modelq}$, and $\text{tr}(\cdot) \in \mathcal{F}_{\modelq}$.

\begin{Def}
	The 
    \emph{inert history}
    $\mathcal{H}_{\modelq}^{\perp \mathcal{F}}$
    of quantum model $\modelq$ is 	the nullspace	
	\begin{align}
		\mathcal{H}_{\modelq}^{\perp \mathcal{F}} \coloneqq \bigl\{ h \in \mathcal{H}_{\modelq}  : f(h) = 0 \text{\emph{ for all }} f \in \mathcal{F}_{\modelq}  \bigr\}.
	\end{align}
\end{Def}

\begin{Def}
	The 
    \emph{inert future}
    $\mathcal{F}_{\modelq}^{\perp \mathcal{H}}$
    of quantum model $\modelq$ is
	the nullspace
	\begin{align}
		\mathcal{F}_{\modelq}^{\perp \mathcal{H}} \coloneqq \bigl\{ f \in \mathcal{F}_{\modelq}  : f(h) = 0 \text{\emph{ for all }} h \in \mathcal{H}_{\modelq} 
		\bigr\}.
	\end{align}
\end{Def}

These kernels, $\mathcal{H}_{\modelq}^{\perp \mathcal{F}}$ and $\mathcal{F}_{\modelq}^{\perp \mathcal{H}}$, can be understood as the respective subspaces of $\mathcal{H}_{\modelq}$ and $\mathcal{F}_{\modelq}$ that 
fail to interact with any element
of the counterpart space. 
These four definitions have obvious analogs for HMMs and GHMMs---namely, the span of history-induced memory states $\mathcal{H}_{\model}$, the span of future-induced linear functionals $\mathcal{F}_{\model}$, and the inert subspaces of these ($\mathcal{H}_{\model}^{\perp \mathcal{F}}$ and $\mathcal{F}_{\model}^{\perp \mathcal{H}}$)~\cite{Uppe97a}.

A classical stochastic process is described fully by conditional probabilities, together with a prior, independent of any generative mechanism. This allows us to leverage history and future spaces to construct generator-independent bases and descriptions of the process. This leads to the notion of a canonical standard GHMM of the process. By identifying a mapping from QHMM to standard GHMM, the classical stochastic processes generated by two different QHMMs can be checked for equivalence by determining if both QHMMs are mapped to the same standard GHMM.

A \emph{wordlist} $\Omega$ is simply a list of words, with elements in $\mathcal{X}^*$.

\begin{Def}
	A history wordlist $\Omega_\text{h}$ is \emph{sufficient} for the process generated by the quantum model $\modelq$ if $\text{\emph{span}} \Bigl\{\text{\emph{span}} \bigl\{ A_w(\sigma) : w \in \Omega_\text{h} \bigr\}
	\cup 	\mathcal{H}_{\modelq}^{\perp \mathcal{F}}	\Bigr\}	 = \mathcal{H}_{\modelq}$.
\end{Def}

\begin{Def}
	A future wordlist $\Omega_\text{f}$ is \emph{sufficient} for the process generated by the quantum model $\modelq$
	if $\text{\emph{span}} \Bigl\{\text{\emph{span}} \bigl\{ \text{\emph{tr}} \bigl[ A_w( \cdot ) \bigr] : w \in \Omega_\text{f} \bigr\}
	\cup \mathcal{F}_{\modelq}^{\perp \mathcal{H}}	\Bigr\}	= \mathcal{F}_{\modelq}$.
\end{Def}

These can analogously be defined for HMMs and GHMMs~\cite{Uppe97a}. The \emph{minimal} sufficient wordlist of a model has size $\ell_\text{min} = \text{dim}(\mathcal{H}_{\model}) - \text{dim}(	\mathcal{H}_{\model}^{\perp \mathcal{F}}) = \text{dim}(\mathcal{F}_{\model}) - \text{dim}(	\mathcal{F}_{\model}^{\perp \mathcal{H}}) $. This minimal size is the same for both history and future wordlists~\cite{Uppe97a}.

For a given classical HMM or GHMM with $|\SSet|$ latent states, the minimal wordlist has no more than $|\SSet|$ words (including the empty word $\varnothing$), since this is guaranteed to be a sufficiently large basis to span the space of latent states~\cite{Uppe97a}. We now provide the analogous statements for QHMMs.

\begin{Prop}
For a given quantum model $\modelq = \bigl(\mathcal{X}, \sigma_0, (A_x)_{x\in\mathcal{X}}\bigr)$, the minimal history and future wordlists each have no more than $d^2$ words (including the empty word $\varnothing$), where $d$ is the dimension of the Hilbert space to which the memory system belongs\footnote{More strictly, $d$ is the dimension of the smallest Hilbert space required to support all reachable memory states $\{\sigma_t\}_{t\in\mathbb{Z}_{\geq0}}$.}.
\end{Prop}

The proof is straightforward; there cannot be more than $d^2$ linearly-independent density matrices acting on a $d$-dimensional Hilbert space, and thus the requisite basis to span the space of latent quantum memory states must be of at most dimension $d^2$.

\begin{Prop}
\label{prop:wordlistlength}
For a given quantum model $\modelq = \bigl(\mathcal{X}, \sigma_0, (A_x)_{x\in\mathcal{X}}\bigr)$, the minimal history and future wordlists each need not contain words longer than $d^2-1$, where $d$ is the dimension of the Hilbert space to which the memory system belongs.
\end{Prop}

This is analogous to Lem.~4.2.14 of Ref.~\cite{Uppe97a}, and follows the same proof, recast in terms of the minimal wordlists for QHMMs rather than GHMMs.

\begin{thm}
\label{thm:identifiability}
	Two quantum models $\modelq = (\mathcal{X},  \sigma_0, (A_x)_{x\in\mathcal{X}})$ and $\modelq' = (\mathcal{X},  \sigma_0', ({A_x}')_{x\in\mathcal{X}})$ generate the same classical stochastic process---i.e., $P_{\modelq}(X_{0:L}) = P_{\modelq'}(X_{0:L})$ for all $L \in \mathbb{N}$---if and only if
    	\begin{enumerate}
		\item 
		$P_{\modelq}( w_f| w_h) =  P_{\modelq'}( w_f| w_h)$  
		for all $w_h \in \pastWL \cup \pastWL'$ and all $w_f \in \futureWL \cup \futureWL'$,
		\item 
		$P_{\modelq}( x w_f| w_h) =  P_{\modelq'}( x w_f|  w_h)$  
		for all $x \in \mathcal{X}$, all $w_h \in \pastWL \cup \pastWL'$, and all $w_f \in \futureWL \cup \futureWL'$, and
		\item 
		$P_{\modelq}( w_f) =  P_{\modelq'}( w_f)$  
		for all $w_f \in \futureWL \cup \futureWL'$
	\end{enumerate}   
	for sufficient history-word lists $\pastWL$ and $\pastWL'$, and sufficient future-word lists $\futureWL$ and $\futureWL'$, respectively.
\end{thm}	

{\bf Proof.} Consider that QHMMs, defined according to Eq.~\eqref{eq:qhmmdef}, are linear models. The latent state space of a QHMM spans the set of density operators acting on a $d$-dimensional Hilbert space, or subset thereof. Recall that this set of density operators corresponds to the set of unit trace, positive-semidefinite, Hermitian operators on the $d$-dimensional Hilbert space, and thus correspond to elements of a $d^2$-dimensional vector space. 
These abstract state vectors can be represented as standard coordinate vectors when the quantum density matrix is 
\emph{vectorized}, e.g., by being brought into the \emph{Liouville space} representation of the quantum state space\footnote{An explicit way to carry out this vectorization is through use of a generalized Bloch representation, which provides a real-valued coordinate vector representation of the quantum state. See App.~\ref{sec:bloch} for details.}. Correspondingly, a QHMM can be cast as a linear model with a latent state space corresponding to elements of an (at most) $d^2$-dimensional vector space. This recasts our QHMM as a valid GHMM; in other words, every QHMM can be represented as a GHMM. Hence, by converting QHMMs $\modelq$ and $\modelq'$ into GHMMs, their equivalence can be checked in the same manner as done for GHMMs. See Thm.~4.3.3 of Ref.~\cite{Uppe97a}, from which the remainder of the proof follows. 

Thus, the identifiability problem for QHMMs is resolved. From this result two useful corollaries readily follow.

\begin{Cor}
\label{cor:identifiabilitylength}
	Two quantum models $\modelq = \bigl(\mathcal{X},  \sigma_0, (A_x)_{x\in\mathcal{X}}\bigr)$ and $\modelq' = \bigl(\mathcal{X},  \sigma_0', ({A_x}')_{x\in\mathcal{X}}\bigr)$ generate the same classical 	stochastic process---i.e., $P_{\modelq}(X_{0:L}) = P_{\modelq'}(X_{0:L})$ for all $L \in \mathbb{N}$---if and only if they agree on the probability all words of length $2d_\text{max}^2-1$, such that 	$P_{\modelq}(X_{0:2 d_\text{max}^2-1}) = P_{\modelq'}(X_{0:2 d_\text{max}^2-1})$, 	where $d_\text{max} = \max\{d, d' \}$.
\end{Cor}

This follows from combining Prop.~\ref{prop:wordlistlength} with Thm.~\ref{thm:identifiability}. All of the probabilities involved in the conditions of Thm.~\ref{thm:identifiability} can be deduced from probabilities of words that need be no longer than one above the length of the concatenation of the longest words in each of $w_h \in \Omega_h \cup \Omega_h'$ and $w_f \in \Omega_f \cup \Omega_f'$, and marginals thereof. Since each QHMM admits minimal wordlists where the elements are each at most of length $d_\text{max}^2-1$, it then follows that if the minimal wordlists are employed for checking the conditions of Thm.~\ref{thm:identifiability}, it is sufficient for all words of length $2d_\text{max}^2-1$ to have identical probabilities for each model. It is also necessary, since if any of these words have differing probabilities, the QHMMs must be generating inequivalent processes.

\begin{Cor}
Thm.~\ref{thm:identifiability} and Cor.~\ref{cor:identifiabilitylength} each still hold when recast in terms of a comparison between a QHMM $\modelq$ and (G)HMM $\model$, with $d'$ replaced by $\sqrt{|\mathcal{S}|}$, where $|\mathcal{S}|$ is of the latent state space dimension of the (G)HMM.
\end{Cor}

This follows since the first step of the proof of Thm.~\ref{thm:identifiability} consists of mapping each QHMM to a GHMM; we need now only do this for the lone QHMM, and compare this to the (G)HMM directly.

\section{Bounding minimal quantum models}
\label{sec:Minimizing}

Just as many different HMMs and GHMMs can generate the same classical stochastic process, so too can the same process be generated by multiple different QHMMs. Practically speaking, the drive for efficiency and parsimony pushes us to opt for the smallest such models. This motivates the question: \emph{What is the minimal quantum model, and what is its size?}

In resolving the identifiability problem we have used that a QHMM can be recast as a GHMM, for which a standard presentation can be constructed. Here, we make further use of this recasting, in order to place lower bounds on the size of the minimal quantum model required to generate a given classical stochastic process.

\begin{thm}
\label{thm:qhmmbound}
The minimal QHMM $\modelq^{\mathrm{min}}$ of a classical stochastic process has a memory dimension of $d_{\mathrm{min}}\geq\sqrt{\ell_{\mathrm{min}}}$, where $\ell_{\mathrm{min}}$ is the length of the minimal history and future wordlists for the process.
\end{thm}
	
{\bf Proof.} Thm.~4.3.10 of Ref.~\cite{Uppe97a}, provides a similar statement with respect to GHMMs. Namely, that the minimal GHMM of a process has $\ell_{\mathrm{min}}$ states. Since a QHMM $\modelq$ of memory dimension $d$ can be recast as a GHMM of size $d^2$, that the minimal GHMM is of dimension $\ell_{\mathrm{min}}$ guarantees that a quantum model cannot exist with dimension smaller than $\sqrt{\ell_{\mathrm{min}}}$.

There are two key respects in which our bound on minimal QHMM size departs from the analogous statement for GHMMs in Thm.~4.3.10 of Ref.~\cite{Uppe97a}. Firstly, that the bound is $\sqrt{\ell_{\mathrm{min}}}$, rather than simply $\ell_{\mathrm{min}}$; and secondly, that it is a bound, rather than a strict equality.

To round out the result, we now provide a systematic approach by which the standard (i.e., minimal) GHMM of a classical stochastic process can be constructed from a quantum model $\modelq$ of the process.

Consider minimal wordlists of the process $\Omega_h$ and $\Omega_f$. These can be directly determined from a modified version of Algorithm 4.2.12 of Ref.~\cite{Uppe97a} adapted for QHMMs, given in App.~\ref{sec:minimalwordlist}.
For each word in the history wordlist, there is a corresponding density matrix that the word $w$ induces: $\sigma^{(w)} = N \bigl[ A_{w}(\sigma_0) \bigr]$.
With $\Omega_h = \{w_\ell\}_{\ell=1}^{\ell_{\mathrm{min}}}$ and 
$\Omega_f = \{w_\ell'\}_{\ell=1}^{\ell_{\mathrm{min}}}$,
we then construct the $HF$ matrix of conditional probabilities that must be shared by all generators of a process given some history and future wordlists:
\begin{align}
HF &= \sum_{\ell=1}^{\ell_{\mathrm{min}}} \sum_{\ell'=1}^{\ell_{\mathrm{min}}}  \ket{\delta_\ell} \! \bra{\delta_{\ell'}} \tr \bigl[ A_{w_{\ell'}'}(\sigma^{(w_\ell)}) \bigr]
~,
\end{align} 
where we have introduced the one-hot row vectors $\bra{\delta_\ell}$ and column vectors $\ket{\delta_\ell}$ 
such that 
$\ket{\delta_\ell} \! \bra{\delta_{\ell'}}$ is an $\ell_{\mathrm{min}}$-by-$\ell_{\mathrm{min}}$ matrix of all zeros except for a single one in the $\ell^\text{th}$ row and ${\ell'}^\text{th}$ column.

The \emph{standard GHMM} of the process can now be written as 
$\model^{\mathrm{std}}=\bigl( \mathcal{X} ,  \boldsymbol{\gamma}_0 , (B^{(x)})_{x \in \mathcal{X}} \bigr)$,
where the row-vector $\boldsymbol{\gamma}_0$
can be obtained from a QHMM via
\begin{align}
\boldsymbol{\gamma}_0 
&=
\sum_{\ell=1}^{\ell_{\mathrm{min}}} 
\tr \bigl[ A_{w_{\ell}'}(\sigma_0) \bigr] \bra{\delta_\ell} (HF)^{-1} 
~,
\end{align}
and the transition matrices can be obtained from a QHMM via
\begin{align}
B^{(x)} 
&= 
\sum_{\ell=1}^{\ell_{\mathrm{min}}} \sum_{\ell'=1}^{\ell_{\mathrm{min}}}  
\tr \bigl[ A_{w_{\ell'}'} \bigl(A_x(\sigma^{(w_\ell)}) \bigr) \bigr]
\ket{\delta_\ell} \! \bra{\delta_{\ell'}} 
(HF)^{-1} 
~.
\end{align}

This definition of the standard GHMM is backwards compatible with the way the standard GHMM has been defined in Ref.~\cite{Uppe97a} for any HMM or GHMM that generates the same process.  Accordingly, any generator of the process---whether classical, quantum, or post quantum---can be identified as producing the same or different classical stochastic process directly via whether they share the same standard GHMM or not.

As should now be familiar, words are generated by the standard GHMM according to 
\begin{align}
P_{\model^{\mathrm{std}}}(w)
=  \boldsymbol{\gamma}_0 B^{(w)}  \boldsymbol{1},
\end{align}
with $B^{(w)} = B^{(x_0)} \cdots B^{(x_{|w|-1})}$. This GHMM spans a latent state space of size $\ell_{\mathrm{\min}}$, and hence corresponds to a minimal GHMM.

In this `standard basis', the column vector
$\boldsymbol{1}$
coincides with the all-ones vector familiar from HMMs.  However, when the minimal GHMM is not an HMM, then  $\boldsymbol{\gamma}_0$ cannot generally be interpreted as a probability distribution over latents.

It should be remarked that, due to the minimal wordlist in general being non-unique, the standard GHMM is also not unique; starting from different minimal wordlists will generally lead to different standard GHMMs. However, if one imposes an ordering on the symbols in the alphabet $\mathcal{X}$, employing the Algorithm of App.~\ref{sec:minimalwordlist} to construct minimal wordlists will then induce unique, ordered minimal history and future wordlists. Correspondingly standard GHMMs constructed in this manner from an ordered event alphabet are unique, and so can be taken as a canonical presentation of the process.

Given that the minimal GHMM has a size equal to that of the minimal wordlist, and our bound for the minimal QHMM size is the square root of this, one may be tempted to imagine this indicates a quadratic quantum advantage. However, given that this is merely a bound on QHMM size, the reality is more subtle.

\begin{thm}
The bound of Thm.~\ref{thm:qhmmbound} on the size of a minimal QHMM of a classical stochastic process is tight for some, but not all, processes.
\end{thm}

{\bf Proof.} We prove this with explicit examples. In App.~\ref{sec:tightexample} we detail a process for which we can provide an explicit QHMM construction that achieves the bound, thus realising a quadratic quantum advantage over the minimal GHMM. That the bound is not always tight is already evidenced by prior results showing that certain processes can be generated by a finite GHMM but no finite QHMM~\cite{Fanizza24_Quantum}. Nevertheless, for completeness, in App.~\ref{sec:looseexample} we detail a process for which the minimal HMM, QHMM, and GHMM all have the same size.

\section{Conclusion}

We have introduced a canonical GHMM representation of QHMMs, such that any generative model---be they classical, quantum, or post-quantum---of a classical stochastic process can be recast in terms of such a canonical representation. With this, we have resolved the general identifiability problem, independent of the physics of the representation; that is, we have shown how to determine whether any two generative models induce the same classical stochastic process. We have further shown that this canonical representation allows us to readily place a lower bound on the dimension of the minimal quantum generative model of any given process.

The bounds on QHMMs are generically quadratically smaller than those of the corresponding GHMMs. Yet, whereas the bound for QHMMs is indeed a bound that is not always tight, the bound for GHMMs is. As we see from our examples, there can be cases where the minimal QHMM is smaller than the minimal GHMM, and there are others where it is not. Yet, it is important to remember that QHMMs are always physically realizable, which is not always the case for a GHMM, and so in practical terms a quantum advantage may still generically remain. In parallel work, we have shown that neural networks leverage the more parsimonious quantum and post-quantum representations of classical stochastic process belief spaces to make efficient use of memory~\cite{Reic24_Neural}. Yet, in this case it is key to note that the number of neurons in the neural network grows only linearly with the number of memory bits, while the state space of a QHMM grows exponentially with the number of memory qubits. This serves as a reminder that the true source of quantum advantage is not because of the presence of amplitudes described by continuous numbers, but rather, from their exponential growth in number of latent dimensions.

We believe there are prospective interesting connections to be made with other fields. The notion of finitely-correlated states~\cite{fannes1992finitely} closely parallel classical stochastic processes, and if one considers a finitely-correlated state to be observed in terms of a fixed basis, they are in effect a model of such a process. Finitely-correlated states are themselves precursors to matrix product states (MPS), and indeed, classical and quantum models of stochastic processes have been related to MPS representations~\cite{monras2010hidden, yang2018matrix, glasser2019expressive, yang2020measures, adhikary2021quantum, yang2025dimension}. GHMMs admit an MPS representation, and we expect a more formal comparison would find that the canonical representation introduced here corresponds to a fixed gauge representation of an infinite MPS. Indeed, the gauge freedom of MPS tensors directly parallels that of GHMM transition matrices. Correspondingly, the bounds determined here could likely also be found from the Liouville-space MPS representation of a QHMM; similarly, related work on bounding QHMM dimension based on spectral properties likely arises from the same underpinning structure~\cite{zonnios2025quantum}.

An exciting extension to be made in future work is to input--output stochastic processes---transducers~\cite{barnett2015computational}---where the observable behavior of the system is conditioned on input stimuli received. A quantum extension of these ideas has been developed, showing memory and thermal advantages analogous to those of quantum models of stochastic processes~\cite{elliott2022quantum, thompson2025energetic}, and GHMM-like transducers have been discussed in Ref.~\cite{Rosas2025ai}. 
By developing a similar notion of input-dependent GHMMs, the canonical representation considered here can be extended to input--output processes, and thus enables us to resolve the identifiability problem and place lower bounds on memory dimension for generation of this richer, more general class of process.

\appendix

\section{Bloch Vectorization}
\label{sec:bloch}

\subsection{Generalized Bloch representation of density matrices}

The state of a qubit $\rho$ can be expressed via its Bloch vector $\vec{a}$: $\rho = I/2 + \vec{a} \cdot \vec{\sigma} / 2$, where  $\vec{\sigma} = (\sigma_x, \sigma_y, \sigma_z)$ is the vector of Pauli matrices.	For a quantum system of arbitrary finite dimension---i.e., a qudit $\rho$ acting on a $d$-dimensional Hilbert space $\mathcal{H}_d$---we can make a generalized Bloch decomposition~\cite{Jako01, Riec24_Ideal}. 

Consider a complete basis $(I/d, \Gamma_1, \Gamma_2, \dots \Gamma_{d^2 - 1})$ for linear operators acting on $\mathcal{H}_d$, such that the Hermitian operators $\Gamma_n$ are all traceless and mutually orthogonal, satisfying
\begin{equation}
		\tr( \Gamma_n) = 0 \; \text{ and } \; \tr(\Gamma_m \Gamma_n) =  \constname \, \delta_{m,n},
\end{equation}
where we set the normalizing constant to be $\constname = \tfrac{d-1}{d}$. Any density matrix then has a unique decomposition in this operator basis, described by the \emph{generalized Bloch vector} $\vec{b} \in \mathbb{R}^{d^2 - 1}$
via 
\begin{align}
	\rho &= \sysref + \vec{b} \cdot \vec{\Gamma} \\
	&= \Bigl( \bigl[ 1 \; \vec{b} \bigr] \otimes I \Bigr) 	\begin{bmatrix}	I/d \\ \Gamma_1 \\ \vdots \\ \Gamma_{d^2 - 1} \end{bmatrix},
	\label{eq:GenBlochExpansion}
\end{align}
where $\vec{\Gamma} = (\Gamma_1, \Gamma_2, \dots \Gamma_{d^2 - 1})$. Density matrices are thus a linear function of 
their $d^2$-dimensional \emph{extended Bloch vector} $\bigl[ 1 \; \vec{b} \bigr] $. This linear function is determined uniquely from the choice of operator basis. Moreover, given any Hermitian operator $M = cI/d + \vec{b} \cdot \vec{\Gamma}$, its extended Bloch vector $\bigl[ c \; \vec{b} \bigr]$ can be obtained via
\begin{align}
	c = \tr(M) \quad \text{and } \quad \vec{b} = \tr(M \vec{\Gamma}) / \constname ~.
\end{align}	

Since the magnitude of the Bloch vector is $b = \sqrt{\frac{\tr(\rho^2) d - 1}{d-1}}$,
the density matrix represents a pure state iff the magnitude of its corresponding Bloch vector is one.  For $d>2$, not all points in the Bloch ball correspond to physical states, but the set of all physical states is nevertheless a convex set---the convex hull of the pure states, and these pure states all lie on a $2(d-1)$-dimensional submanifold of the $(d^{2}-2)$-dimensional surface of the Bloch sphere~\cite{Jako01}.

\subsection{Generalized Bloch representation of QHMMs}

The superoperators $\{A_x\}$ of a QHMM correspond to completely positive (CP) maps on the quantum memory system. Marginalizing over the output events, the overall channel is CP, and also trace preserving (CPTP). Each of the $|\mathcal{X}|$ subchannels described by the superoperators can be fully determined, using linearity, from measuring the output of $d^2$ independent inputs $\{ \rho_{(n)} \}_{n=1}^{d^2}$ to each subchannel. Moreover, each subchannel has a generalized Bloch representation that can be readily constructed from such experiments.

Each subchannel $A_x$ is a linear superoperator acting on a density matrix, which is in turn a linear function of its extended Bloch vector.  Accordingly, each subchannel can be represented as a linear operator $G^{(x)}$ acting on the extended Bloch vector:
\begin{align}
\bigl[ 1 \; \vec{b}_n \bigr] G^{(x)} = \bigl[ c_n \; \vec{b}_n' \bigr],
\end{align}	
where $\bigl[ 1 \; \vec{b}_n \bigr]$ is the extended Bloch vector of the subchannel input state $\rho_{(n)}$, and $ \bigl[ c_n \; \vec{b}_n' \bigr] $ is the extended Bloch vector of $A_x(\rho_{(n)})$, with $c_n = \tr \bigl[ A_x(\rho_{(n)}) \bigr]$ and $\vec{b}_n' =  \tr \bigl[ A_x(\rho_{(n)}) \vec{\Gamma} \bigr] / \constname$.

Stacking these equations for $d^2$ linearly independent inputs from the memory system state space, i.e.,
\begin{align}
	\underbrace{
	\begin{bmatrix}
		1 & \vec{b}_1 \\
		1 & \vec{b}_2 \\
		\vdots & \vdots \\
		1 & \vec{b}_{d^2} 
	\end{bmatrix}	
}_{\eqqcolon \mathfrak{B}}
G^{(x)}
&= 
	\underbrace{
	\begin{bmatrix}
		c_1 & \vec{b}_1' \\
		c_2 & \vec{b}_2' \\
		\vdots & \vdots \\
		c_{d^2} & \vec{b}_{d^2}' 
	\end{bmatrix}	
}_{\eqqcolon \mathfrak{B}'},
\end{align}	
and designating new extended Bloch matrices $\mathfrak{B}$ and $\mathfrak{B}'$, allows us to directly construct $G^{(x)}$: 
\begin{align}
G^{(x)} = \mathfrak{B}^{-1} \mathfrak{B}'.
\end{align}	
Note that $\mathfrak{B}$ is always invertible, since we have insisted on $d^2$ linearly independent input states.

This construction directly yields a $d^2$-dimensional GHMM representation of the stochastic process. We obtain the net transition operator $G = \sum_{x \in \mathcal{X}} G^{(x)}$, and the stationary vectors of the process $\stationary = \stationary G = \bigl[ 1 \;  \vec{b}_{\stationary} \bigr]$ and $\ones = G \ones = \bigl[ 1 \; 0 \, \dots \, 0 \bigr]^\top$. 

If we insist on a GHMM representation with an all-ones vector, then we must perform a similarity transform:
\begin{align}
	G^{(x)} \mapsto S G^{(x)} S^{-1},
\end{align}	
with $S \bigl[ 1 \; 0 \, \dots \, 0 \bigr]^\top = \bigl[ 1 \; 1 \, \dots \, 1 \bigr]^\top$.

\subsection{Liouville space representation}

An alternative, more direct approach to vectorization is simply to `reshape' density operators into vectors and superoperators into matrices. That is, suppose our QHMM has superoperators $\{A_x\}_x$ described by a set of Kraus operators $\{K_{xy}\}_{x,y}$. A GHMM can be constructed with transition operators $\{G^{(x)}\}_{x \in \mathcal{X}} = \{\sum_{y\in\mathcal{Y}}K_{xy} \otimes K_{xy}^*\}_{x \in \mathcal{X}}$, the standard Liouville-space operators. In place of density matrices, quantum states are represented by complex-valued column vectors: $\rho = \sum_n p_n \ket{\psi_n} \! \bra{\psi_n} \mapsto \sum_n p_n \ket{\psi_n} \otimes \ket{\psi_n}^*$~\cite{Gyamfi2020fundamentals}.

We can transpose everything to obtain the typical GHMM format\footnote{If we wish to have an all-ones vector, then we must again also perform a corresponding similarity transformation.}. It is worth noting that these different representations will all have the same spectral properties on their non-zero eigenspaces. 

\section{Constructing minimal wordlists from QHMMs}
\label{sec:minimalwordlist}

Here we provide an adaptation of Algorithm 4.2.12 of Ref.~\cite{Uppe97a} for constructing minimal wordlists from GHMMs to allow for a minimal wordlist to be constructed directly from a QHMM.

\begin{algorithm}[H]
\caption{\textsf{Sufficient history wordlist from QHMM}}
\begin{flushleft}
\emph{Inputs}: QHMM $\modelq = \bigl(\mathcal{X}, \sigma_0, (A_x)_{x\in\mathcal{X}}\bigr)$. \\
\emph{Outputs}: Sufficient history wordlist $\Omega_h$.
\end{flushleft}
\begin{algorithmic}[1]
\STATE Let $Q$ be a queue (first in, first out list) of words, initially containing only the empty word $\varnothing$. Let $\Omega_h$ be a word list, initially empty (not containing the empty word). 
\STATE Let $z$ be the word at the tail of $Q$. Test whether or not $A_z(\sigma_0)$ lies within $\mathrm{span} \bigl\{ A_w(\sigma_0) : w \in \Omega_h \bigr\}.$ If yes, discard $z$ and proceed to Step 4. Otherwise, add $z$ to the tail of $\Omega_h$ and proceed to Step 3.
\STATE For each $x\in\mathcal{X}$, add the word $zx$ to the head of $Q$.
\STATE If $Q$ is empty, return to Step 2. Otherwise STOP.
\end{algorithmic}
\end{algorithm}

This will construct a sufficent history wordlist for the QHMM. A similar algorithm can be employed to construct a sufficient future wordlist $\Omega_f$ by replacing the $A_w(\sigma_0)$ with the functionals $\mathrm{tr}[A_w(.)]$. If $\mathcal{X}$ is an ordered set, then this will construct an ordered wordlist. 

A minimal wordlist can then be found by pruning the relevant null space. For history and future matrices $H$ and $F$ constructed from wordlists $\Omega_h$ and $\Omega_f$ respectively, determine $F(H)$. Remove columns from this matrix one-by-one until all columns are linearly dependent. Then, remove the words from $\Omega_h$ corresponding to the deleted columns in order to obtain a minimal history wordlist $\Omega_h'$.

\section{Examples}

\subsection{Example: Tightness of bound}
\label{sec:tightexample}

\begin{figure}
	\includegraphics[width=0.75\columnwidth]{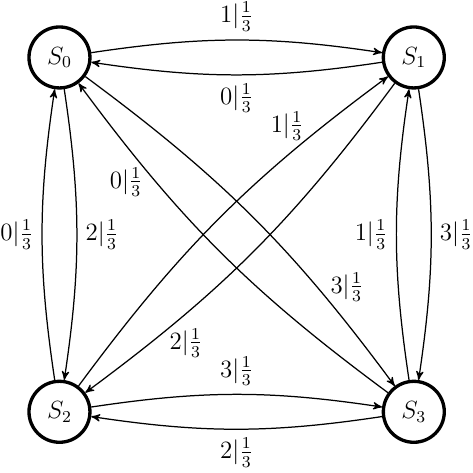}
	\caption{HMM representation of the example process for App.~\ref{sec:tightexample}. Nodes represent latent states of the model. The edge notation $x|p$ denotes that the indicated transition between latent states takes place with probability $p$, and corresponds to event $x$ occuring.}
	\label{fig:tightexample}
\end{figure}

For the first example, consider the process depicted by the HMM in Fig.~\ref{fig:tightexample}. This is inspired by the communication task described in Ref.~\cite{elliott2025strict}. We will show for this example that the minimal wordlist contains four words, whilst also prescribing an explicit QHMM construction of dimension two. This thus substantiates the tightness of our bound in Thm.~\ref{thm:qhmmbound}, and the quantum advantage.

Let us begin with an explicit QHMM construction. Consider a unitary operator $U$ defined to act as follows:
\begin{align}
U\ket{\psi_0}\ket{0}&=\frac{1}{\sqrt{3}}(\ket{\psi_1}\ket{1}+\ket{\psi_2}\ket{2}+\ket{\psi_3}\ket{3})\nonumber\\
U\ket{\psi_1}\ket{0}&=\frac{1}{\sqrt{3}}(\ket{\psi_0}\ket{0}+\ket{\psi_2}\ket{2}+e^{\frac{i\pi}{3}}\ket{\psi_3}\ket{3})\nonumber\\
U\ket{\psi_2}\ket{0}&=\frac{1}{\sqrt{3}}(-\ket{\psi_0}\ket{0}+\ket{\psi_1}\ket{1}+e^{\frac{-i\pi}{3}}\ket{\psi_3}\ket{3})\nonumber\\
U\ket{\psi_3}\ket{0}&=\frac{1}{\sqrt{3}}(-e^{\frac{-i\pi}{3}}\ket{\psi_0}\ket{0}+\ket{\psi_1}\ket{1}+e^{\frac{i\pi}{3}}\ket{\psi_2}\ket{2}),
\label{eq:exampleunitary}
\end{align}
where the first subsystem corresponds to the quantum memory system, the second the ancilla for the output event label, and there is no trashed ancillary subsystem. The quantum memory states $\{\ket{\psi_0},\ket{\psi_1},\ket{\psi_2},\ket{\psi_3}\}$ are in one-to-one correspondence with the latent states of the HMM.

It can readily be verified that these dynamics give rise to the correct probabilities that generate the process. An explicit representation of $U$ in terms of a given basis can be obtained using the method of Refs.~\cite{binder2018practical} and ~\cite{liu2019optimal}. From this, we can then readily deduce Kraus operators and corresponding superoperators to define the QHMM proper.

From Eq.~(\ref{eq:exampleunitary}) it can be observed that the quantum memory states satisfy $\ket{\psi_2}=\ket{\psi_0}-\ket{\psi_1}$ and $\ket{\psi_3}=\ket{\psi_0}-e^{-i\pi/3}\ket{\psi_1}$. That is, two of the states can be obtained as linear combinations of the other two. Thus, all four states can be embedded within a two-dimensional Hilbert space, and thus, the QHMM has only a two-dimensional memory system.

From the HMM depicted, it is straightforward to see that a sufficient history wordlist is given by $\{0,1,2,3\}$, since the process is Markovian, with all events mapping to a different state. The associated history matrix is the $4\times4$ identity matrix. Meanwhile, a sufficient future wordlist can be shown to be $\{\varnothing, 0, 1, 2\}$. Since the history matrix is the identity matrix, this must also be a minimal wordlist. Thus, the minimal GHMM has dimension four. Thence, the QHMM saturates the bound (i.e., has dimension given by the square root of the size of the minimal wordlist), and has a quadratic advantage over the minimal GHMM and HMM.

\subsection{Example: No quantum advantage}
\label{sec:looseexample}

\begin{figure}[tb]
\includegraphics[width=0.75\columnwidth]{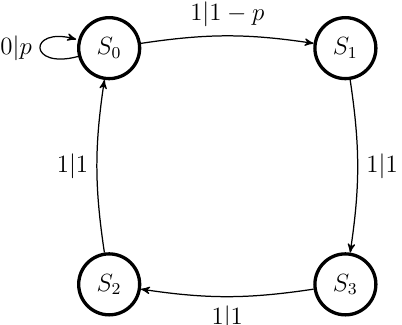}
	\caption{HMM representation of the example process for App.~\ref{sec:looseexample}. Nodes represent latent states of the model. The edge notation $x|p$ denotes that the indicated transition between latent states takes place with probability $p$, and corresponds to event $x$ occurring.}
	\label{fig:looseexample}
\end{figure}

For the second example, consider the process depicted in Fig.~\ref{fig:looseexample}. This process is described by a four state HMM, and thus by construction we therefore know that the minimal GHMM and QHMM of this process are both also of at most dimension four.

This HMM is both unifilar (i.e., if the current latent state and next output symbol are known, then the next latent state is known with certainty, such that 
the conditional Shannon entropy of the next hidden state satisfies
$\text{H}(S_1|S_0,X_0)=0$) and co-unifilar (meaning that if the current latent state and most recent past symbol are known, then the previous latent state is known with certainty; $\text{H}(S_0|S_1,X_0)=0$). It is known that a model satisfying this property satisfies $\text{H}(S)=\text{I}(\past{X};\fut{X})$---i.e., the entropy of the latent states then equals the mutual information between past and future---and has the minimal Shannon entropy across all classical and quantum models~\cite{crutchfield2009time, elliott2021memory}.

The steady-state probabilities of the HMM latent states are $\stationary=[1, 1-p, 1-p, 1-p]/(4-3p)$. For $p\ll1$, the Shannon entropy of this exceeds $\log_2(3)$, meaning that this information cannot be encoded in fewer than four dimensions. Since an explicit construction exists with four states, this means that the minimal HMMs and QHMMs that can generate this process have four dimensions.

But what is the size of the minimal wordlist? Taking the HMM and setting $[1, 0, 0, 0]$ as the initial state, Algorithm 4.2.12 of Ref.~\cite{Uppe97a} can readily be applied to determine a sufficient history wordlist can be given by $\{\varnothing, 1, 11, 111\}$. Meanwhile, a sufficient future wordlist is given by $\{\varnothing, 0, 10, 110\}$. Since the associated history matrix is the identity, the minimal wordlists must contain four words, and hence the minimal GHMM is also of dimension four.

\acknowledgements

We thank our many colleagues for insightful discussions over the years that have shaped our thoughts on this line of research. TJE is supported by the University of Manchester Dame Kathleen Ollerenshaw Fellowship.

\bibliography{chaos,ref}

\begin{thebibliography}{38}%
\makeatletter
\providecommand \@ifxundefined [1]{%
 \@ifx{#1\undefined}
}%
\providecommand \@ifnum [1]{%
 \ifnum #1\expandafter \@firstoftwo
 \else \expandafter \@secondoftwo
 \fi
}%
\providecommand \@ifx [1]{%
 \ifx #1\expandafter \@firstoftwo
 \else \expandafter \@secondoftwo
 \fi
}%
\providecommand \natexlab [1]{#1}%
\providecommand \enquote  [1]{``#1''}%
\providecommand \bibnamefont  [1]{#1}%
\providecommand \bibfnamefont [1]{#1}%
\providecommand \citenamefont [1]{#1}%
\providecommand \href@noop [0]{\@secondoftwo}%
\providecommand \href [0]{\begingroup \@sanitize@url \@href}%
\providecommand \@href[1]{\@@startlink{#1}\@@href}%
\providecommand \@@href[1]{\endgroup#1\@@endlink}%
\providecommand \@sanitize@url [0]{\catcode `\\12\catcode `\$12\catcode
  `\&12\catcode `\#12\catcode `\^12\catcode `\_12\catcode `\%12\relax}%
\providecommand \@@startlink[1]{}%
\providecommand \@@endlink[0]{}%
\providecommand \url  [0]{\begingroup\@sanitize@url \@url }%
\providecommand \@url [1]{\endgroup\@href {#1}{\urlprefix }}%
\providecommand \urlprefix  [0]{URL }%
\providecommand \Eprint [0]{\href }%
\providecommand \doibase [0]{https://doi.org/}%
\providecommand \selectlanguage [0]{\@gobble}%
\providecommand \bibinfo  [0]{\@secondoftwo}%
\providecommand \bibfield  [0]{\@secondoftwo}%
\providecommand \translation [1]{[#1]}%
\providecommand \BibitemOpen [0]{}%
\providecommand \bibitemStop [0]{}%
\providecommand \bibitemNoStop [0]{.\EOS\space}%
\providecommand \EOS [0]{\spacefactor3000\relax}%
\providecommand \BibitemShut  [1]{\csname bibitem#1\endcsname}%
\let\auto@bib@innerbib\@empty
\bibitem [{\citenamefont {Burnham}\ and\ \citenamefont
  {Anderson}(2003)}]{burnham2003model}%
  \BibitemOpen
  \bibfield  {author} {\bibinfo {author} {\bibfnamefont {K.~P.}\ \bibnamefont
  {Burnham}}\ and\ \bibinfo {author} {\bibfnamefont {D.~R.}\ \bibnamefont
  {Anderson}},\ }\href@noop {} {\emph {\bibinfo {title} {Model Selection and
  Multimodel Inference: A Practical Information-Theoretic Approach}}}\
  (\bibinfo  {publisher} {Springer New York},\ \bibinfo {year}
  {2003})\BibitemShut {NoStop}%
\bibitem [{\citenamefont {Crutchfield}(1994)}]{Crut92c}%
  \BibitemOpen
  \bibfield  {author} {\bibinfo {author} {\bibfnamefont {J.~P.}\ \bibnamefont
  {Crutchfield}},\ }\bibfield  {title} {\bibinfo {title} {The calculi of
  emergence: Computation, dynamics, and induction},\ }\href@noop {} {\bibfield
  {journal} {\bibinfo  {journal} {Physica D}\ }\textbf {\bibinfo {volume}
  {75}},\ \bibinfo {pages} {11} (\bibinfo {year} {1994})}\BibitemShut {NoStop}%
\bibitem [{\citenamefont {Gu}\ \emph {et~al.}(2012)\citenamefont {Gu},
  \citenamefont {Wiesner}, \citenamefont {Rieper},\ and\ \citenamefont
  {Vedral}}]{Gu12a}%
  \BibitemOpen
  \bibfield  {author} {\bibinfo {author} {\bibfnamefont {M.}~\bibnamefont
  {Gu}}, \bibinfo {author} {\bibfnamefont {K.}~\bibnamefont {Wiesner}},
  \bibinfo {author} {\bibfnamefont {E.}~\bibnamefont {Rieper}},\ and\ \bibinfo
  {author} {\bibfnamefont {V.}~\bibnamefont {Vedral}},\ }\bibfield  {title}
  {\bibinfo {title} {Quantum mechanics can reduce the complexity of classical
  models},\ }\href {https://doi.org/10.1038/ncomms1761} {\bibfield  {journal}
  {\bibinfo  {journal} {Nature Comm.}\ }\textbf {\bibinfo {volume} {3}},\
  \bibinfo {pages} {1} (\bibinfo {year} {2012})}\BibitemShut {NoStop}%
\bibitem [{\citenamefont {Elliott}(2021)}]{elliott2021memory}%
  \BibitemOpen
  \bibfield  {author} {\bibinfo {author} {\bibfnamefont {T.~J.}\ \bibnamefont
  {Elliott}},\ }\bibfield  {title} {\bibinfo {title} {Memory compression and
  thermal efficiency of quantum implementations of nondeterministic hidden
  {M}arkov models},\ }\href@noop {} {\bibfield  {journal} {\bibinfo  {journal}
  {Physical Review A}\ }\textbf {\bibinfo {volume} {103}},\ \bibinfo {pages}
  {052615} (\bibinfo {year} {2021})}\BibitemShut {NoStop}%
\bibitem [{\citenamefont {Monràs}\ and\ \citenamefont
  {Winter}(2016)}]{Monras16_Quantum}%
  \BibitemOpen
  \bibfield  {author} {\bibinfo {author} {\bibfnamefont {A.}~\bibnamefont
  {Monràs}}\ and\ \bibinfo {author} {\bibfnamefont {A.}~\bibnamefont
  {Winter}},\ }\bibfield  {title} {\bibinfo {title} {Quantum learning of
  classical stochastic processes: {T}he completely positive realization
  problem},\ }\href {https://doi.org/10.1063/1.4936935} {\bibfield  {journal}
  {\bibinfo  {journal} {Journal of Mathematical Physics}\ }\textbf {\bibinfo
  {volume} {57}},\ \bibinfo {pages} {015219} (\bibinfo {year} {2016})},\
  \Eprint
  {https://arxiv.org/abs/https://pubs.aip.org/aip/jmp/article-pdf/doi/10.1063/1.4936935/11140332/015219\_1\_online.pdf}
  {https://pubs.aip.org/aip/jmp/article-pdf/doi/10.1063/1.4936935/11140332/015219\_1\_online.pdf}
  \BibitemShut {NoStop}%
\bibitem [{\citenamefont {Blackwell}\ and\ \citenamefont
  {Koopmans}(1957)}]{Blac57a}%
  \BibitemOpen
  \bibfield  {author} {\bibinfo {author} {\bibfnamefont {D.}~\bibnamefont
  {Blackwell}}\ and\ \bibinfo {author} {\bibfnamefont {L.}~\bibnamefont
  {Koopmans}},\ }\bibfield  {title} {\bibinfo {title} {On the identifiability
  problem for functions of {Markov} chains},\ }\href@noop {} {\bibfield
  {journal} {\bibinfo  {journal} {Ann. Math. Statist.}\ }\textbf {\bibinfo
  {volume} {28}},\ \bibinfo {pages} {1011} (\bibinfo {year}
  {1957})}\BibitemShut {NoStop}%
\bibitem [{\citenamefont {Upper}(1997)}]{Uppe97a}%
  \BibitemOpen
  \bibfield  {author} {\bibinfo {author} {\bibfnamefont {D.~R.}\ \bibnamefont
  {Upper}},\ }\emph {\bibinfo {title} {Theory and Algorithms for Hidden
  {M}arkov Models and Generalized Hidden {M}arkov Models}},\ \href@noop {}
  {Ph.D. thesis},\ \bibinfo  {school} {University of California}, \bibinfo
  {address} {Berkeley} (\bibinfo {year} {1997}),\ \bibinfo {note} {{P}ublished
  by University Microfilms Intl, Ann Arbor, Michigan}\BibitemShut {NoStop}%
\bibitem [{\citenamefont
  {Khintchine}(1934)}]{khintchine1934korrelationstheorie}%
  \BibitemOpen
  \bibfield  {author} {\bibinfo {author} {\bibfnamefont {A.}~\bibnamefont
  {Khintchine}},\ }\bibfield  {title} {\bibinfo {title} {Korrelationstheorie
  der station{\"a}ren stochastischen {P}rozesse},\ }\href@noop {} {\bibfield
  {journal} {\bibinfo  {journal} {Mathematische Annalen}\ }\textbf {\bibinfo
  {volume} {109}},\ \bibinfo {pages} {604} (\bibinfo {year}
  {1934})}\BibitemShut {NoStop}%
\bibitem [{\citenamefont {Kolmogorov}(2018)}]{kolmogorov2018foundations}%
  \BibitemOpen
  \bibfield  {author} {\bibinfo {author} {\bibfnamefont {A.~N.}\ \bibnamefont
  {Kolmogorov}},\ }\href@noop {} {\emph {\bibinfo {title} {Foundations of the
  theory of probability: {S}econd English Edition}}}\ (\bibinfo  {publisher}
  {Courier Dover Publications},\ \bibinfo {year} {2018})\BibitemShut {NoStop}%
\bibitem [{\citenamefont {Garner}(2015)}]{garner2015phase}%
  \BibitemOpen
  \bibfield  {author} {\bibinfo {author} {\bibfnamefont {A.}~\bibnamefont
  {Garner}},\ }\emph {\bibinfo {title} {Phase and interference phenomena in
  generalised probabilistic theories}},\ \href@noop {} {Ph.D. thesis},\
  \bibinfo  {school} {Oxford University, UK} (\bibinfo {year}
  {2015})\BibitemShut {NoStop}%
\bibitem [{\citenamefont {Fanizza}\ \emph {et~al.}(2024)\citenamefont
  {Fanizza}, \citenamefont {Lumbreras},\ and\ \citenamefont
  {Winter}}]{Fanizza24_Quantum}%
  \BibitemOpen
  \bibfield  {author} {\bibinfo {author} {\bibfnamefont {M.}~\bibnamefont
  {Fanizza}}, \bibinfo {author} {\bibfnamefont {J.}~\bibnamefont {Lumbreras}},\
  and\ \bibinfo {author} {\bibfnamefont {A.}~\bibnamefont {Winter}},\
  }\bibfield  {title} {\bibinfo {title} {Quantum theory in finite dimension
  cannot explain every general process with finite memory},\ }\href@noop {}
  {\bibfield  {journal} {\bibinfo  {journal} {Communications in Mathematical
  Physics}\ }\textbf {\bibinfo {volume} {405}},\ \bibinfo {pages} {50}
  (\bibinfo {year} {2024})}\BibitemShut {NoStop}%
\bibitem [{\citenamefont {Balle}\ \emph {et~al.}(2015)\citenamefont {Balle},
  \citenamefont {Panangaden},\ and\ \citenamefont
  {Precup}}]{Balle15_Canonical}%
  \BibitemOpen
  \bibfield  {author} {\bibinfo {author} {\bibfnamefont {B.}~\bibnamefont
  {Balle}}, \bibinfo {author} {\bibfnamefont {P.}~\bibnamefont {Panangaden}},\
  and\ \bibinfo {author} {\bibfnamefont {D.}~\bibnamefont {Precup}},\
  }\bibfield  {title} {\bibinfo {title} {A canonical form for weighted automata
  and applications to approximate minimization},\ }in\ \href@noop {} {\emph
  {\bibinfo {booktitle} {2015 30th Annual ACM/IEEE Symposium on Logic in
  Computer Science}}}\ (\bibinfo {organization} {IEEE},\ \bibinfo {year}
  {2015})\ pp.\ \bibinfo {pages} {701--712}\BibitemShut {NoStop}%
\bibitem [{\citenamefont {Shalizi}\ and\ \citenamefont
  {Crutchfield}(2001)}]{Shal98a}%
  \BibitemOpen
  \bibfield  {author} {\bibinfo {author} {\bibfnamefont {C.~R.}\ \bibnamefont
  {Shalizi}}\ and\ \bibinfo {author} {\bibfnamefont {J.~P.}\ \bibnamefont
  {Crutchfield}},\ }\bibfield  {title} {\bibinfo {title} {Computational
  mechanics: Pattern and prediction, structure and simplicity},\ }\href@noop {}
  {\bibfield  {journal} {\bibinfo  {journal} {J. Stat. Phys.}\ }\textbf
  {\bibinfo {volume} {104}},\ \bibinfo {pages} {817} (\bibinfo {year}
  {2001})}\BibitemShut {NoStop}%
\bibitem [{\citenamefont {Monras}\ \emph {et~al.}(2010)\citenamefont {Monras},
  \citenamefont {Beige},\ and\ \citenamefont {Wiesner}}]{monras2010hidden}%
  \BibitemOpen
  \bibfield  {author} {\bibinfo {author} {\bibfnamefont {A.}~\bibnamefont
  {Monras}}, \bibinfo {author} {\bibfnamefont {A.}~\bibnamefont {Beige}},\ and\
  \bibinfo {author} {\bibfnamefont {K.}~\bibnamefont {Wiesner}},\ }\bibfield
  {title} {\bibinfo {title} {Hidden quantum {M}arkov models and non-adaptive
  read-out of many-body states},\ }\href@noop {} {\bibfield  {journal}
  {\bibinfo  {journal} {arXiv preprint arXiv:1002.2337}\ } (\bibinfo {year}
  {2010})}\BibitemShut {NoStop}%
\bibitem [{\citenamefont {Glasser}\ \emph {et~al.}(2019)\citenamefont
  {Glasser}, \citenamefont {Sweke}, \citenamefont {Pancotti}, \citenamefont
  {Eisert},\ and\ \citenamefont {Cirac}}]{glasser2019expressive}%
  \BibitemOpen
  \bibfield  {author} {\bibinfo {author} {\bibfnamefont {I.}~\bibnamefont
  {Glasser}}, \bibinfo {author} {\bibfnamefont {R.}~\bibnamefont {Sweke}},
  \bibinfo {author} {\bibfnamefont {N.}~\bibnamefont {Pancotti}}, \bibinfo
  {author} {\bibfnamefont {J.}~\bibnamefont {Eisert}},\ and\ \bibinfo {author}
  {\bibfnamefont {I.}~\bibnamefont {Cirac}},\ }\bibfield  {title} {\bibinfo
  {title} {Expressive power of tensor-network factorizations for probabilistic
  modeling},\ }\href@noop {} {\bibfield  {journal} {\bibinfo  {journal}
  {Advances in neural information processing systems}\ }\textbf {\bibinfo
  {volume} {32}} (\bibinfo {year} {2019})}\BibitemShut {NoStop}%
\bibitem [{\citenamefont {Binder}\ \emph {et~al.}(2018)\citenamefont {Binder},
  \citenamefont {Thompson},\ and\ \citenamefont {Gu}}]{binder2018practical}%
  \BibitemOpen
  \bibfield  {author} {\bibinfo {author} {\bibfnamefont {F.~C.}\ \bibnamefont
  {Binder}}, \bibinfo {author} {\bibfnamefont {J.}~\bibnamefont {Thompson}},\
  and\ \bibinfo {author} {\bibfnamefont {M.}~\bibnamefont {Gu}},\ }\bibfield
  {title} {\bibinfo {title} {Practical unitary simulator for non-{M}arkovian
  complex processes},\ }\href@noop {} {\bibfield  {journal} {\bibinfo
  {journal} {Physical Review Letters}\ }\textbf {\bibinfo {volume} {120}},\
  \bibinfo {pages} {240502} (\bibinfo {year} {2018})}\BibitemShut {NoStop}%
\bibitem [{\citenamefont {Liu}\ \emph {et~al.}(2019)\citenamefont {Liu},
  \citenamefont {Elliott}, \citenamefont {Binder}, \citenamefont {Di~Franco},\
  and\ \citenamefont {Gu}}]{liu2019optimal}%
  \BibitemOpen
  \bibfield  {author} {\bibinfo {author} {\bibfnamefont {Q.}~\bibnamefont
  {Liu}}, \bibinfo {author} {\bibfnamefont {T.~J.}\ \bibnamefont {Elliott}},
  \bibinfo {author} {\bibfnamefont {F.~C.}\ \bibnamefont {Binder}}, \bibinfo
  {author} {\bibfnamefont {C.}~\bibnamefont {Di~Franco}},\ and\ \bibinfo
  {author} {\bibfnamefont {M.}~\bibnamefont {Gu}},\ }\bibfield  {title}
  {\bibinfo {title} {Optimal stochastic modeling with unitary quantum
  dynamics},\ }\href@noop {} {\bibfield  {journal} {\bibinfo  {journal}
  {Physical Review A}\ }\textbf {\bibinfo {volume} {99}},\ \bibinfo {pages}
  {062110} (\bibinfo {year} {2019})}\BibitemShut {NoStop}%
\bibitem [{\citenamefont {Adhikary}\ \emph {et~al.}(2020)\citenamefont
  {Adhikary}, \citenamefont {Srinivasan}, \citenamefont {Gordon},\ and\
  \citenamefont {Boots}}]{adhikary2020expressiveness}%
  \BibitemOpen
  \bibfield  {author} {\bibinfo {author} {\bibfnamefont {S.}~\bibnamefont
  {Adhikary}}, \bibinfo {author} {\bibfnamefont {S.}~\bibnamefont
  {Srinivasan}}, \bibinfo {author} {\bibfnamefont {G.}~\bibnamefont {Gordon}},\
  and\ \bibinfo {author} {\bibfnamefont {B.}~\bibnamefont {Boots}},\ }\bibfield
   {title} {\bibinfo {title} {Expressiveness and learning of hidden quantum
  {M}arkov models},\ }in\ \href@noop {} {\emph {\bibinfo {booktitle}
  {International Conference on Artificial Intelligence and Statistics}}}\
  (\bibinfo {organization} {PMLR},\ \bibinfo {year} {2020})\ pp.\ \bibinfo
  {pages} {4151--4161}\BibitemShut {NoStop}%
\bibitem [{\citenamefont {Loomis}\ and\ \citenamefont
  {Crutchfield}(2019)}]{loomis2019strong}%
  \BibitemOpen
  \bibfield  {author} {\bibinfo {author} {\bibfnamefont {S.~P.}\ \bibnamefont
  {Loomis}}\ and\ \bibinfo {author} {\bibfnamefont {J.~P.}\ \bibnamefont
  {Crutchfield}},\ }\bibfield  {title} {\bibinfo {title} {Strong and weak
  optimizations in classical and quantum models of stochastic processes},\
  }\href@noop {} {\bibfield  {journal} {\bibinfo  {journal} {Journal of
  Statistical Physics}\ }\textbf {\bibinfo {volume} {176}},\ \bibinfo {pages}
  {1317} (\bibinfo {year} {2019})}\BibitemShut {NoStop}%
\bibitem [{\citenamefont {Ghafari}\ \emph {et~al.}(2019)\citenamefont
  {Ghafari}, \citenamefont {Tischler}, \citenamefont {Thompson}, \citenamefont
  {Gu}, \citenamefont {Shalm}, \citenamefont {Verma}, \citenamefont {Nam},
  \citenamefont {Patel}, \citenamefont {Wiseman},\ and\ \citenamefont
  {Pryde}}]{ghafari2019dimensional}%
  \BibitemOpen
  \bibfield  {author} {\bibinfo {author} {\bibfnamefont {F.}~\bibnamefont
  {Ghafari}}, \bibinfo {author} {\bibfnamefont {N.}~\bibnamefont {Tischler}},
  \bibinfo {author} {\bibfnamefont {J.}~\bibnamefont {Thompson}}, \bibinfo
  {author} {\bibfnamefont {M.}~\bibnamefont {Gu}}, \bibinfo {author}
  {\bibfnamefont {L.~K.}\ \bibnamefont {Shalm}}, \bibinfo {author}
  {\bibfnamefont {V.~B.}\ \bibnamefont {Verma}}, \bibinfo {author}
  {\bibfnamefont {S.~W.}\ \bibnamefont {Nam}}, \bibinfo {author} {\bibfnamefont
  {R.~B.}\ \bibnamefont {Patel}}, \bibinfo {author} {\bibfnamefont {H.~M.}\
  \bibnamefont {Wiseman}},\ and\ \bibinfo {author} {\bibfnamefont {G.~J.}\
  \bibnamefont {Pryde}},\ }\bibfield  {title} {\bibinfo {title} {Dimensional
  quantum memory advantage in the simulation of stochastic processes},\
  }\href@noop {} {\bibfield  {journal} {\bibinfo  {journal} {Physical Review
  X}\ }\textbf {\bibinfo {volume} {9}},\ \bibinfo {pages} {041013} (\bibinfo
  {year} {2019})}\BibitemShut {NoStop}%
\bibitem [{\citenamefont {Elliott}\ \emph {et~al.}(2020)\citenamefont
  {Elliott}, \citenamefont {Yang}, \citenamefont {Binder}, \citenamefont
  {Garner}, \citenamefont {Thompson},\ and\ \citenamefont
  {Gu}}]{elliott2020extreme}%
  \BibitemOpen
  \bibfield  {author} {\bibinfo {author} {\bibfnamefont {T.~J.}\ \bibnamefont
  {Elliott}}, \bibinfo {author} {\bibfnamefont {C.}~\bibnamefont {Yang}},
  \bibinfo {author} {\bibfnamefont {F.~C.}\ \bibnamefont {Binder}}, \bibinfo
  {author} {\bibfnamefont {A.~J.~P.}\ \bibnamefont {Garner}}, \bibinfo {author}
  {\bibfnamefont {J.}~\bibnamefont {Thompson}},\ and\ \bibinfo {author}
  {\bibfnamefont {M.}~\bibnamefont {Gu}},\ }\bibfield  {title} {\bibinfo
  {title} {Extreme dimensionality reduction with quantum modeling},\
  }\href@noop {} {\bibfield  {journal} {\bibinfo  {journal} {Physical Review
  Letters}\ }\textbf {\bibinfo {volume} {125}},\ \bibinfo {pages} {260501}
  (\bibinfo {year} {2020})}\BibitemShut {NoStop}%
\bibitem [{\citenamefont {Zonnios}\ \emph {et~al.}(2025)\citenamefont
  {Zonnios}, \citenamefont {Boyd},\ and\ \citenamefont
  {Binder}}]{zonnios2025quantum}%
  \BibitemOpen
  \bibfield  {author} {\bibinfo {author} {\bibfnamefont {M.}~\bibnamefont
  {Zonnios}}, \bibinfo {author} {\bibfnamefont {A.}~\bibnamefont {Boyd}},\ and\
  \bibinfo {author} {\bibfnamefont {F.}~\bibnamefont {Binder}},\ }\bibfield
  {title} {\bibinfo {title} {Quantum generation of stochastic processes:
  spectral invariants and memory bounds},\ }\href@noop {} {\bibfield  {journal}
  {\bibinfo  {journal} {New Journal of Physics}\ } (\bibinfo {year}
  {2025})}\BibitemShut {NoStop}%
\bibitem [{\citenamefont {Loomis}\ and\ \citenamefont
  {Crutchfield}(2020)}]{loomis2020thermal}%
  \BibitemOpen
  \bibfield  {author} {\bibinfo {author} {\bibfnamefont {S.~P.}\ \bibnamefont
  {Loomis}}\ and\ \bibinfo {author} {\bibfnamefont {J.~P.}\ \bibnamefont
  {Crutchfield}},\ }\bibfield  {title} {\bibinfo {title} {Thermal efficiency of
  quantum memory compression},\ }\href@noop {} {\bibfield  {journal} {\bibinfo
  {journal} {Physical Review Letters}\ }\textbf {\bibinfo {volume} {125}},\
  \bibinfo {pages} {020601} (\bibinfo {year} {2020})}\BibitemShut {NoStop}%
\bibitem [{\citenamefont {Riechers}\ \emph {et~al.}(2025)\citenamefont
  {Riechers}, \citenamefont {Elliott},\ and\ \citenamefont
  {Shai}}]{Reic24_Neural}%
  \BibitemOpen
  \bibfield  {author} {\bibinfo {author} {\bibfnamefont {P.~M.}\ \bibnamefont
  {Riechers}}, \bibinfo {author} {\bibfnamefont {T.~J.}\ \bibnamefont
  {Elliott}},\ and\ \bibinfo {author} {\bibfnamefont {A.~S.}\ \bibnamefont
  {Shai}},\ }\bibfield  {title} {\bibinfo {title} {Neural networks leverage
  nominally quantum and post-quantum representations},\ }\href@noop {}
  {\bibfield  {journal} {\bibinfo  {journal} {arXiv:2507.07432}\ } (\bibinfo
  {year} {2025})}\BibitemShut {NoStop}%
\bibitem [{\citenamefont {Fannes}\ \emph {et~al.}(1992)\citenamefont {Fannes},
  \citenamefont {Nachtergaele},\ and\ \citenamefont
  {Werner}}]{fannes1992finitely}%
  \BibitemOpen
  \bibfield  {author} {\bibinfo {author} {\bibfnamefont {M.}~\bibnamefont
  {Fannes}}, \bibinfo {author} {\bibfnamefont {B.}~\bibnamefont
  {Nachtergaele}},\ and\ \bibinfo {author} {\bibfnamefont {R.~F.}\ \bibnamefont
  {Werner}},\ }\bibfield  {title} {\bibinfo {title} {Finitely correlated states
  on quantum spin chains},\ }\href@noop {} {\bibfield  {journal} {\bibinfo
  {journal} {Communications in mathematical physics}\ }\textbf {\bibinfo
  {volume} {144}},\ \bibinfo {pages} {443} (\bibinfo {year}
  {1992})}\BibitemShut {NoStop}%
\bibitem [{\citenamefont {Yang}\ \emph {et~al.}(2018)\citenamefont {Yang},
  \citenamefont {Binder}, \citenamefont {Narasimhachar},\ and\ \citenamefont
  {Gu}}]{yang2018matrix}%
  \BibitemOpen
  \bibfield  {author} {\bibinfo {author} {\bibfnamefont {C.}~\bibnamefont
  {Yang}}, \bibinfo {author} {\bibfnamefont {F.~C.}\ \bibnamefont {Binder}},
  \bibinfo {author} {\bibfnamefont {V.}~\bibnamefont {Narasimhachar}},\ and\
  \bibinfo {author} {\bibfnamefont {M.}~\bibnamefont {Gu}},\ }\bibfield
  {title} {\bibinfo {title} {Matrix product states for quantum stochastic
  modelling},\ }\href@noop {} {\bibfield  {journal} {\bibinfo  {journal}
  {Physical Review Letters}\ }\textbf {\bibinfo {volume} {121}},\ \bibinfo
  {pages} {260602} (\bibinfo {year} {2018})}\BibitemShut {NoStop}%
\bibitem [{\citenamefont {Yang}\ \emph {et~al.}(2020)\citenamefont {Yang},
  \citenamefont {Binder}, \citenamefont {Gu},\ and\ \citenamefont
  {Elliott}}]{yang2020measures}%
  \BibitemOpen
  \bibfield  {author} {\bibinfo {author} {\bibfnamefont {C.}~\bibnamefont
  {Yang}}, \bibinfo {author} {\bibfnamefont {F.~C.}\ \bibnamefont {Binder}},
  \bibinfo {author} {\bibfnamefont {M.}~\bibnamefont {Gu}},\ and\ \bibinfo
  {author} {\bibfnamefont {T.~J.}\ \bibnamefont {Elliott}},\ }\bibfield
  {title} {\bibinfo {title} {Measures of distinguishability between stochastic
  processes},\ }\href@noop {} {\bibfield  {journal} {\bibinfo  {journal}
  {Physical Review E}\ }\textbf {\bibinfo {volume} {101}},\ \bibinfo {pages}
  {062137} (\bibinfo {year} {2020})}\BibitemShut {NoStop}%
\bibitem [{\citenamefont {Adhikary}\ \emph {et~al.}(2021)\citenamefont
  {Adhikary}, \citenamefont {Srinivasan}, \citenamefont {Miller}, \citenamefont
  {Rabusseau},\ and\ \citenamefont {Boots}}]{adhikary2021quantum}%
  \BibitemOpen
  \bibfield  {author} {\bibinfo {author} {\bibfnamefont {S.}~\bibnamefont
  {Adhikary}}, \bibinfo {author} {\bibfnamefont {S.}~\bibnamefont
  {Srinivasan}}, \bibinfo {author} {\bibfnamefont {J.}~\bibnamefont {Miller}},
  \bibinfo {author} {\bibfnamefont {G.}~\bibnamefont {Rabusseau}},\ and\
  \bibinfo {author} {\bibfnamefont {B.}~\bibnamefont {Boots}},\ }\bibfield
  {title} {\bibinfo {title} {Quantum tensor networks, stochastic processes, and
  weighted automata},\ }in\ \href@noop {} {\emph {\bibinfo {booktitle}
  {International Conference on Artificial Intelligence and Statistics}}}\
  (\bibinfo {organization} {PMLR},\ \bibinfo {year} {2021})\ pp.\ \bibinfo
  {pages} {2080--2088}\BibitemShut {NoStop}%
\bibitem [{\citenamefont {Yang}\ \emph {et~al.}(2025)\citenamefont {Yang},
  \citenamefont {Florido-Llin{\`a}s}, \citenamefont {Gu},\ and\ \citenamefont
  {Elliott}}]{yang2025dimension}%
  \BibitemOpen
  \bibfield  {author} {\bibinfo {author} {\bibfnamefont {C.}~\bibnamefont
  {Yang}}, \bibinfo {author} {\bibfnamefont {M.}~\bibnamefont
  {Florido-Llin{\`a}s}}, \bibinfo {author} {\bibfnamefont {M.}~\bibnamefont
  {Gu}},\ and\ \bibinfo {author} {\bibfnamefont {T.~J.}\ \bibnamefont
  {Elliott}},\ }\bibfield  {title} {\bibinfo {title} {Dimension reduction in
  quantum sampling of stochastic processes},\ }\href@noop {} {\bibfield
  {journal} {\bibinfo  {journal} {npj Quantum Information}\ }\textbf {\bibinfo
  {volume} {11}},\ \bibinfo {pages} {34} (\bibinfo {year} {2025})}\BibitemShut
  {NoStop}%
\bibitem [{\citenamefont {Barnett}\ and\ \citenamefont
  {Crutchfield}(2015)}]{barnett2015computational}%
  \BibitemOpen
  \bibfield  {author} {\bibinfo {author} {\bibfnamefont {N.}~\bibnamefont
  {Barnett}}\ and\ \bibinfo {author} {\bibfnamefont {J.~P.}\ \bibnamefont
  {Crutchfield}},\ }\bibfield  {title} {\bibinfo {title} {Computational
  mechanics of input--output processes: Structured transformations and the
  $\varepsilon$-transducer},\ }\href@noop {} {\bibfield  {journal} {\bibinfo
  {journal} {Journal of Statistical Physics}\ }\textbf {\bibinfo {volume}
  {161}},\ \bibinfo {pages} {404} (\bibinfo {year} {2015})}\BibitemShut
  {NoStop}%
\bibitem [{\citenamefont {Elliott}\ \emph {et~al.}(2022)\citenamefont
  {Elliott}, \citenamefont {Gu}, \citenamefont {Garner},\ and\ \citenamefont
  {Thompson}}]{elliott2022quantum}%
  \BibitemOpen
  \bibfield  {author} {\bibinfo {author} {\bibfnamefont {T.~J.}\ \bibnamefont
  {Elliott}}, \bibinfo {author} {\bibfnamefont {M.}~\bibnamefont {Gu}},
  \bibinfo {author} {\bibfnamefont {A.~J.}\ \bibnamefont {Garner}},\ and\
  \bibinfo {author} {\bibfnamefont {J.}~\bibnamefont {Thompson}},\ }\bibfield
  {title} {\bibinfo {title} {Quantum adaptive agents with efficient long-term
  memories},\ }\href@noop {} {\bibfield  {journal} {\bibinfo  {journal}
  {Physical Review X}\ }\textbf {\bibinfo {volume} {12}},\ \bibinfo {pages}
  {011007} (\bibinfo {year} {2022})}\BibitemShut {NoStop}%
\bibitem [{\citenamefont {Thompson}\ \emph {et~al.}(2025)\citenamefont
  {Thompson}, \citenamefont {Riechers}, \citenamefont {Garner}, \citenamefont
  {Elliott},\ and\ \citenamefont {Gu}}]{thompson2025energetic}%
  \BibitemOpen
  \bibfield  {author} {\bibinfo {author} {\bibfnamefont {J.}~\bibnamefont
  {Thompson}}, \bibinfo {author} {\bibfnamefont {P.~M.}\ \bibnamefont
  {Riechers}}, \bibinfo {author} {\bibfnamefont {A.~J.}\ \bibnamefont
  {Garner}}, \bibinfo {author} {\bibfnamefont {T.~J.}\ \bibnamefont
  {Elliott}},\ and\ \bibinfo {author} {\bibfnamefont {M.}~\bibnamefont {Gu}},\
  }\bibfield  {title} {\bibinfo {title} {Energetic advantages for quantum
  agents in online execution of complex strategies},\ }\href@noop {} {\bibfield
   {journal} {\bibinfo  {journal} {arXiv preprint arXiv:2503.19896}\ }
  (\bibinfo {year} {2025})}\BibitemShut {NoStop}%
\bibitem [{\citenamefont {Rosas}\ \emph {et~al.}(2025)\citenamefont {Rosas},
  \citenamefont {Boyd},\ and\ \citenamefont {Baltieri}}]{Rosas2025ai}%
  \BibitemOpen
  \bibfield  {author} {\bibinfo {author} {\bibfnamefont {F.}~\bibnamefont
  {Rosas}}, \bibinfo {author} {\bibfnamefont {A.}~\bibnamefont {Boyd}},\ and\
  \bibinfo {author} {\bibfnamefont {M.}~\bibnamefont {Baltieri}},\ }\bibfield
  {title} {\bibinfo {title} {Ai in a vat: {F}undamental limits of efficient
  world modelling for agent sandboxing and interpretability},\ }\href@noop {}
  {\bibfield  {journal} {\bibinfo  {journal} {arXiv preprint arXiv:2504.04608}\
  } (\bibinfo {year} {2025})}\BibitemShut {NoStop}%
\bibitem [{\citenamefont {Jak{\'o}bczyk}\ and\ \citenamefont
  {Siennicki}(2001)}]{Jako01}%
  \BibitemOpen
  \bibfield  {author} {\bibinfo {author} {\bibfnamefont {L.}~\bibnamefont
  {Jak{\'o}bczyk}}\ and\ \bibinfo {author} {\bibfnamefont {M.}~\bibnamefont
  {Siennicki}},\ }\bibfield  {title} {\bibinfo {title} {Geometry of {B}loch
  vectors in two-qubit system},\ }\href@noop {} {\bibfield  {journal} {\bibinfo
   {journal} {Physics Letters A}\ }\textbf {\bibinfo {volume} {286}},\ \bibinfo
  {pages} {383} (\bibinfo {year} {2001})}\BibitemShut {NoStop}%
\bibitem [{\citenamefont {Riechers}\ \emph {et~al.}(2024)\citenamefont
  {Riechers}, \citenamefont {Gupta}, \citenamefont {Kolchinsky},\ and\
  \citenamefont {Gu}}]{Riec24_Ideal}%
  \BibitemOpen
  \bibfield  {author} {\bibinfo {author} {\bibfnamefont {P.~M.}\ \bibnamefont
  {Riechers}}, \bibinfo {author} {\bibfnamefont {C.}~\bibnamefont {Gupta}},
  \bibinfo {author} {\bibfnamefont {A.}~\bibnamefont {Kolchinsky}},\ and\
  \bibinfo {author} {\bibfnamefont {M.}~\bibnamefont {Gu}},\ }\bibfield
  {title} {\bibinfo {title} {Thermodynamically ideal quantum state inputs to
  any device},\ }\href {https://doi.org/10.1103/PRXQuantum.5.030318} {\bibfield
   {journal} {\bibinfo  {journal} {PRX Quantum}\ }\textbf {\bibinfo {volume}
  {5}},\ \bibinfo {pages} {030318} (\bibinfo {year} {2024})}\BibitemShut
  {NoStop}%
\bibitem [{\citenamefont {Gyamfi}(2020)}]{Gyamfi2020fundamentals}%
  \BibitemOpen
  \bibfield  {author} {\bibinfo {author} {\bibfnamefont {J.~A.}\ \bibnamefont
  {Gyamfi}},\ }\bibfield  {title} {\bibinfo {title} {Fundamentals of quantum
  mechanics in {L}iouville space},\ }\href@noop {} {\bibfield  {journal}
  {\bibinfo  {journal} {European Journal of Physics}\ }\textbf {\bibinfo
  {volume} {41}},\ \bibinfo {pages} {063002} (\bibinfo {year}
  {2020})}\BibitemShut {NoStop}%
\bibitem [{\citenamefont {Elliott}(2025)}]{elliott2025strict}%
  \BibitemOpen
  \bibfield  {author} {\bibinfo {author} {\bibfnamefont {T.~J.}\ \bibnamefont
  {Elliott}},\ }\bibfield  {title} {\bibinfo {title} {Strict advantage of
  complex quantum theory in a communication task},\ }\href@noop {} {\bibfield
  {journal} {\bibinfo  {journal} {Physical Review A}\ }\textbf {\bibinfo
  {volume} {111}},\ \bibinfo {pages} {062401} (\bibinfo {year}
  {2025})}\BibitemShut {NoStop}%
\bibitem [{\citenamefont {Crutchfield}\ \emph {et~al.}(2009)\citenamefont
  {Crutchfield}, \citenamefont {Ellison},\ and\ \citenamefont
  {Mahoney}}]{crutchfield2009time}%
  \BibitemOpen
  \bibfield  {author} {\bibinfo {author} {\bibfnamefont {J.~P.}\ \bibnamefont
  {Crutchfield}}, \bibinfo {author} {\bibfnamefont {C.~J.}\ \bibnamefont
  {Ellison}},\ and\ \bibinfo {author} {\bibfnamefont {J.~R.}\ \bibnamefont
  {Mahoney}},\ }\bibfield  {title} {\bibinfo {title} {Time’s barbed arrow:
  {I}rreversibility, crypticity, and stored information},\ }\href@noop {}
  {\bibfield  {journal} {\bibinfo  {journal} {Physical Review Letters}\
  }\textbf {\bibinfo {volume} {103}},\ \bibinfo {pages} {094101} (\bibinfo
  {year} {2009})}\BibitemShut {NoStop}%
\end{thebibliography}%
\end{document}